\documentclass[11pt]{article}
\usepackage[margin=1in]{geometry}

\usepackage{amsmath}
\usepackage{amsfonts}
\usepackage{graphicx}   
\usepackage{multirow}    
\usepackage{makecell}
\usepackage{caption}
\usepackage{subcaption}
\usepackage{setspace}
\usepackage{url}
\usepackage{times}
\usepackage{bm}
\usepackage{natbib}
\usepackage{xcolor}
\usepackage{hyperref}
\hypersetup{
    colorlinks,
    linkcolor={red!70!black},
    citecolor={blue!80!black},
    urlcolor={blue!80!black}
}

\newcommand{\1}{1}
\newcommand{\bA}{A_\cD}
\newcommand{\ba}{a_\cD}

\newcommand{\bbeta}{\beta}
\newcommand{\bX}{X_\cD} 
\newcommand{\bx}{x_\cD}   
\newcommand{\bY}{Y_\cD}

\newcommand{\cD}{\mathcal{D}}

\newcommand{\bXcN}{\bX}

\newcommand{\tl}{{\log( \tau - 1/d )}}
\newcommand{\eps}{\epsilon}

\newcommand{\tA}{\tilde{A}} 
\newcommand{\ta}{\tilde{a}}

\newcommand{\trips}[1]{{\left\vert\kern-0.25ex\left\vert\kern-0.25ex\left\vert #1
        \right\vert\kern-0.25ex\right\vert\kern-0.25ex\right\vert}}

\newcommand{\ind}{\mathrel{\text{\scalebox{1.07}{$\perp\mkern-10mu\perp$}}}}

\newtheorem{assumption}{Assumption}\newtheorem{theorem}{Theorem}\newtheorem{corollary}{Corollary}\newtheorem{proof}{Proof}\newtheorem{example}{Example}

\DeclareMathOperator{\E}{E}
\DeclareMathOperator{\Var}{Var}
\DeclareMathOperator{\spl}{spl}
\DeclareMathOperator{\Prob}{pr}
\DeclareMathOperator{\dd}{~d}
\DeclareMathOperator{\Top}{T}
\DeclareMathOperator{\logit}{logit}
\setcitestyle{authoryear, open={(},close={)}}

\begin{document}

\title{\textbf{Generalized propensity score approach to causal inference with spatial interference}}

\author{
  \textsc{By A. B. Giffin, B. J. Reich, S. Yang}\\
  \emph{Department of Statistics, North Carolina State University} \\
  giffin.andrew@gmail.com, bjreich@ncsu.edu, syang24@ncsu.edu
  \and
  \textsc{A. G. Rappold}\\
  \emph{Environmental Protection Agency}\\
  rappold.ana@epa.gov
}

\maketitle

\begin{abstract}
\noindent
Many spatial phenomena exhibit treatment interference where treatments at one location may affect the response at other locations. Because interference violates the stable unit treatment value assumption, standard methods for causal inference do not apply. We propose a new causal framework to recover direct and spill-over effects in the presence of spatial interference, taking into account that treatments at nearby locations are more influential than treatments at locations further apart. Under the no unmeasured confounding assumption, we show that a generalized propensity score is sufficient to remove all measured confounding. To reduce dimensionality issues, we propose a Bayesian spline-based regression model accounting for a sufficient set of variables for the generalized propensity score. A simulation study demonstrates the accuracy and coverage properties. We apply the method to estimate the causal effect of wildland fires on air pollution in the Western United States over 2005--2018.\\

\end{abstract}

\section{Introduction}\label{s:intro}

Understanding spatial processes in the environmental and health sciences has taken on new importance as we grapple with emerging ecological and epidemiological issues. Much of the research in these areas are associative in nature despite the effects of interests being causal \citep{bind2019causal}. This is a result of both the frequent necessity of using observational data, but also the difficulty of implementing causal inference tools on data that exhibit spatial dependence and, in particular, interference. Interference is the phenomenon in which treatments at one location may affect the response at other locations. Naturally, with spatially-dependent processes, a treatment may impact the response nearby, leading to interference. 

An example of spatial interference is the relationship between wildland fires and air pollution. Treating wildland fires as the treatment and pollution as the response, it is clear that the treatment can substantially impact the response at the location of treatment and at distant locations. In this example all available data are observational, and therefore isolating average causal treatment effects requires accounting for confounding variables. Even in the ideal case where all potential confounders are observed across locations, it is unclear how to condition on these confounders without knowing their specific spatial relationships with the treatment and response. Conditioning on confounders at all locations, which is one way around this, is impractical for all but the smallest studies.

The difficulty that arises from interference in the context of spatially dependant processes is immediately apparent from the vantage of the potential outcomes framework developed by \cite{rubin1974estimating}. For a binary treatment without interference, there are two unit-level potential outcomes to consider. Under general treatment interference, there are $2^n$ unit-level potential outcomes to consider, where $n$ is the total number of units, because each treatment permutation across all units represents a distinct treatment. In the case of  geostatistical models that contain uncountably many spatial locations, the problem becomes even more intractable. For this reason, beginning with \cite{cox1958planning} much of the causal inference literature assumes away interference. The no-interference assumption is now usually invoked as one-half of the ubiquitous stable unit treatment value assumption \citep{rubin1980randomization}. 

Relaxations to the no-interference assumption generally involve placing assumptions on the form of interference. Partial interference, a term coined by \cite{sobel2006randomized}, was the first relaxation developed, specifically for modeling vaccination treatments which are known to induce herd immunity.  This assumption defines disjoint groups or clusters a priori which may exhibit interference, but precludes interference between groups. This form of interference was originally considered with experimental data by  \cite{halloran1991study, halloran1995causal}, but expanded to non-randomized data by \cite{hudgens2008toward, tchetgen2012causal, liu2014large, papadogeorgou2019causal}. The dual nature of this form of interference allows for information on both the direct treatment effects as well as the indirect or spill-over effects from interference. Additionally, the deluge of network data has resulted in a literature which allows for interference along edges of a pre-specified graph \citep{athey2018exact}. 

Spatially indexed data have been analyzed using both the partial interference and network interference strategies. For naturally clustered spatial data, the partial interference assumption can be used, e.g., as in \cite{perez2014assessing} and \cite{zigler2012estimating}. Spatial data can also be simplified to the network setting. For areal data, this often entails creating a graph with edges between neighboring units, as in \cite{verbitsky2012causal}. This, however, discards information about the distance between units. 

Despite these advances, there has been little exploration of strictly spatial assumptions on the form of interference. To fill this gap in the literature, we propose a new framework to recover causal direct and spill-over effects in the presence of spatial interference, while taking into account the high dimensionality of the problem. We develop a generalized propensity score to account for spatial dependence in the distribution of treatment. To further reduce the size of the problem, we propose a model which accounts for a sufficient set of summary variables rather than the full generalized propensity score itself.

The proposed approach has a number of advantages over using a partial interference or network interference assumption. The partial interference assumption is only reasonable for limited cases when the data naturally cluster a significant distance apart. Moreover, the partial interference grouping must be specified a priori. The network interference assumption, while more flexible, abandons key spatial information about the distance between points, which may be crucial in the presence of true spatial confounding.  Our proposed method retains all spatial information, and allows for the kernel range to be estimated concurrently.

\section{Potential outcomes, interference, and identification}\label{s:prelims} 

Assume that data are available at $n$ spatial locations $s \in \{s_1, \ldots, s_n\} \subset \cD \subset \mathbb{R}^2$. For spatial location $s$ define $X_s \in \mathbb{R}^p$ as the relevant covariates and  $Y_s \in \mathbb{R}^1$ the response. We will consider both real-valued and binary treatments $A_s$. We use subscript $_\cD$ to refer to the full fields of random variables, e.g., 
$\bX = \{X_{s}: s \in \cD \}$. Variables with subscript $_{-s}$ denote all locations in $\cD$ excluding $s$. Lowercase letters refer to realizations of the variables.

Without restrictions, the response $Y_s$ is potentially a function of $X_\cD$ and $A_\cD$ at all locations, greatly increasing the number of potential outcomes. To make this manageable while still taking spatial interference into account, we assume that the potential outcome $Y_s(\ba)$ depends on treatment field $\ba$ through two mechanisms; a direct treatment, $a_s$, and an indirect/spill-over treatment, $\ta_{\tau,s} = \int_{\cD \backslash s} \omega_\tau \left( \| s - s' \| \right) a_{s'}~\text{d}s'$, where $\omega_\tau(\cdot) : \mathbb{R}^+ \mapsto [0,1]$ is a kernel function with bandwidth $\tau>0$. This constitutes a general class of interference structures. Examples \ref{ex:PI} and \ref{ex:GK} provide two important cases.

\begin{example}\label{ex:PI}
For clustered data, $\omega_\tau(d) = I(d < \tau)$ implies partial interference when the clusters are smaller than $\tau$ in diameter and separated by at least  $\tau$. Here the potential outcome exhibits stratified interference or anonymous interaction \citep{manski2013identification}; i.e., $Y_s(\ba)$ depends on its own treatment and the aggregate treatment of other locations in its cluster.
\end{example}

\begin{example} \label{ex:GK}
  When $\omega_\tau (d) = \exp \{ - ( d /\tau )^2 \}$ takes this Gaussian kernel form with bandwidth $\tau$, interference decays smoothly over space. 
\end{example}

Because only finitely many locations are observed in practice, the integral form of $\ta_{\tau,s}$ must be approximated with a sum. One approach is to assume that  $\{a_1,\ldots,a_n\}$ are the average treatments over $n$ regions that partition $\cD$. This is a tractable approach that is particularly useful for binary treatments. Another more general approach is to treat $a_s$ as a smooth function that can be well approximated by summing over $n$ locations. In this paper we focus on the former, and approximate $\ta_{\tau,s}$ with the form $\ta_{\tau,s} = \sum_{s' \in \{s_1, \ldots, s_n\} \backslash s} \omega_\tau \left( \| s - s' \| \right) a_{s'}$.

Implicitly, we assume that for any $s$ and treatments $\ba$ and $\ba'$, $Y_s(\ba) = Y_s(\ba')$ if $a_s = a_s'$ and $\ta_s = \ta_s'$.  This simplified treatment allows us to parsimoniously define the individual potential outcomes for all possible treatment fields $\ba$ in terms of only the local direct and spill-over treatments: $Y_s( a_s,~ \ta_{\tau,s})$.

Identification of the treatments effects follows from the following assumptions:

\begin{assumption}[Unconfoundedness]
\label{unconfoundedness}
  For all $\ba$, $Y_s(\ba) = Y_s( a_s, \ta_{\tau,s}) \ind \bA \mid  \bXcN$.
\end{assumption}

\begin{assumption}[Positivity]
For all $\bx$ with $\Prob(\bX = \bx) > 0$, 
$\Prob(\bA = \ba \mid  \bX = \bx) > 0$ for all $\ba$. \label{positivity}
\end{assumption}

\begin{assumption}[Consistency]
	The potential outcome $Y_s(a_s, \ta_{\tau,s}) = Y_s$ when $A_s = a_s$ and $\tA_s = \ta_{\tau,s}$.
\label{consistency}
\end{assumption}

For finite $\cD$, with only the assumptions above, treatments effects theoretically are identifiable. However, identification requires the number of repeated field observations to be at least $2^n$, which is rare. To make the situation tractable, we make two additional assumptions about our data as follows:

\begin{assumption}[Marginal Structural Model]
The potential outcomes model take the form
\begin{align}
  Y_s\left( a_s,~ \ta_{\tau,s} \right) 
  &~=~ \beta_0 + \delta_1 a_s + \delta_2 \ta_{\tau,s} + h(\bXcN) + \epsilon_s,\label{eq:responseForm}
\end{align} where $h(\bX)$ is a general function of $\bX$, and $e_s$ is an error process that is independent of $A_\cD$ and $X_\cD$. Here $\delta_1$ and $\delta_2$ quantify the direct and spill-over effects of treatment, respectively; $\tau$ quantifies the range of the spill-over effect $\ta$.
\label{a:responseForm}
\end{assumption}

Under Assumptions \ref{unconfoundedness}--\ref{a:responseForm}, (\ref{eq:responseForm}) is identifiable in the sense that
\begin{align}
  \E \left\{Y_s(\ba)~\mid ~\bX\right\} = 
  \E \left\{Y_s(\ba)~\mid ~\bX, A_s= a_s, \tA = \ta \right\} =
  \E \left( Y_s~\mid ~\bX, A_s= a_s, \tA = \ta \right). \label{identificationFormula}
\end{align}
The first equality follows from Assumptions \ref{unconfoundedness} and \ref{positivity}. The second follows from Assumptions \ref{consistency} and \ref{a:responseForm}. 

It is instructive to consider the dependence that is created by these assumptions. $\bX$ is unrestricted, and is therefore plausibly spatially correlated. Because the direct treatment mechanism is a function of $\bX$, $A_\cD$ will likely  reflect any spatial structure in $X_\cD$. $\bY$ may reflect both general spatial dependence from $\bX$ as well as any induced spatial dependence from $\bA$.

\section{The generalized propensity score is a balancing score}\label{s:unconfoundedness}

The identification formula (\ref{identificationFormula}) implies that we can estimate $\delta_1$, $\delta_2$, and $\tau$ using the regression model
\begin{align*}
Y_s = \delta_1 A_s + \delta_2 \tA_{\tau,s} + h(\bX)_s + \eps_s,
\end{align*}
if $h(\cdot)$ is known and $\eps_s$ is a mean zero error process.  In most cases, though,  $h(\bX)$ is not known. The standard causal inference strategy at this point is to condition on $\bX$ itself, if known. However, even when $\bX$ is known, in the context of spatial analysis it is high dimensional. Specifically, for unit $s$ it does not suffice to condition on $X_s$, but rather requires conditioning on $X$ at all locations. With both high-dimensional confounders as well as our assumptions about the treatment mechanism, the natural path forward is to condition on the propensity of treatment \citep{rosenbaum1983central}.  

In a setting without interference, and thus only direct treatment effects, the standard propensity score $e_s$ for binary treatments is defined as $e_s(\bX) = P(A_s=1 \mid \bX)$. This is easily extended to real valued treatments using the form $e_s(\bX) = f(A_s = \eta \mid \bX), \eta \in \mathbb{R}$. In both cases, $e_s$ simply summarizes the conditional distribution of treatment. The propensity score is an example of a balancing score: a function of the covariates that, once conditioned on, induces independence between the treatment and covariates. If all confounders are included in $X$, then $e_s$, rather than $\bX$, may be conditioned on for unbiased treatment effects. When $\bX$ is high-dimensional, as in our motivating example, this is a substantial dimension reduction.

Under interference, with treatment components $a_s$ and $\ta_{\tau,s}$, the propensity score approach can still be utilized, by defining the propensity of treatment to be a summary of the conditional distribution of $(A_s, \tA_s)$.  To this end, we define $g_{\tau,s}$ to be the joint propensity of $A_s$ and $\tA_{\tau,s}$:
\begin{align}
	g_{\tau,s}(\bXcN) 
	= f ( A_s = \eta , ~\tA_{\tau,s} = \nu \mid \bXcN ), \quad \eta, \nu \in \mathbb{R}.
 \label{eq:g}
\end{align}
We refer to the bivariate density function $g_{\tau,s}$ as the generalized propensity score. Importantly, this general form of $g_{\tau,s}$ allows for treatments $A_\cD$ to be correlated, which in turn may cause dependence between $A_s$ and $\tA_{\tau,s}$. 

The key insight is that $g_{\tau,s}$ is a balancing score. This implies that, paired with our no unmeasured confounders assumption, the observed treatments and potential outcomes are independent conditional on $g_{\tau,s}$. This is the strategy which we use to recover unbiased estimates of our key coefficients $\delta_1$ and $\delta_2$. Theorem \ref{GPStheorem} shows this formally, by extending the analogous result for propensity scores for continuous treatments by \cite{hirano2004propensity} to our generalized propensity score $g_{\tau,s}$.

\begin{theorem}[$g_{\tau,s}$ is a balancing score]
 \emph{Given Assumptions \ref{unconfoundedness}--\ref{a:responseForm}, then for all locations $s$ and spill-over treatment levels $\nu$,} 
\begin{align*} Y_s( \ba) = Y_s( a_s, \ta_{\tau,s} = \nu) \ind (
  A_s, \tA_{\tau,s}) \mid  g_{\tau,s}(\bXcN).
\end{align*}
\label{GPStheorem}
\end{theorem}
The proof is provided in the Appendix \ref{a:proof}.

By Theorem \ref{GPStheorem} it suffices to adjust for $g_{\tau,s}$ to remove confounding bias. Namely, Theorem \ref{GPStheorem} implies that
\begin{align*}
  \E \left\{ Y_s(\ba) ~\mid ~ g_{\tau,s}(\bX) \right\}
  = \E \left\{ Y_s ~\mid ~ g_{\tau,s}(\bX), A_s = a_s, \tA_{\tau,s} = \ta_{\tau,s}\right\}.
\end{align*}
This suggests that we can adjust for potential confounding by incorporating $g_{\tau,s}$ into the regression model.

\section{Modeling the generalized propensity score}\label{s:modelingTheGPS}

Estimating $g_{\tau,s}$ is difficult.  It is a bivariate distribution function over $\bX$, and non-parametric estimation of even univariate density functions suffers from dimensionality issues. To overcome this, we make the following dimension reduction assumption.
 
\begin{assumption}[$g_{\tau,s}$ is a parametric distribution]  \label{parametricDensity}  
\emph{$g_{\tau,s}$ is a bivariate parametric density with parameters $\bar{Z}_s = (Z_s^{(1)}, \ldots, Z_s^{(K)})$ that are a functions of $\tau$ and $\bXcN$.   }
\end{assumption}
That is, the distribution of $(A,\tA_{\tau,s})$ can be completely summarized by low-dimensional parameters $\bar{Z}_s$. 

\begin{example}
If $\bA$ are 
independent and Gaussian then $\tA_{\tau,s}$ is itself Gaussian. Setting $Z_s^1, \ldots, Z_s^4$ to be the mean and variance of both $A_s$ and $\tA_{\tau,s}$ completely summarizes its distribution.
\end{example}

\begin{corollary}
  Given Assumptions \ref{unconfoundedness}--\ref{parametricDensity}, then for all locations $s$
\begin{align*}
  Y_s( \ba) = Y_s( a_s, \ta_{\tau,s}) \ind ( A_s, \tA_{\tau,s} ) \mid  \bar{Z}_s.
\end{align*}
\end{corollary}
This follows immediately from Theorem 1.

This states that conditioning on $\bar{Z}_s$ is equivalent to conditioning directly on the distribution $g_{\tau,s}$, and so our Theorem 1 result of unconfoundedness given $g_{\tau}$  extends to the considerably more tractable situation of unconfoundedness given $\bar{Z}$.  Identification of $\delta_1$ and $\delta_2$ follows from the conditional independence in Corollary 1, as shown in (\ref{fullIdentification}). Equation (\ref{splineIdentification}) sketches the manner in which the components of $g_{\tau,s}$ will be conditioned on using B-splines. Let $^*$ denote true values; variables without $^*$ being estimated values. Based on
\begin{align}
  \begin{split}
  \E\left\{ Y_s(\ba) \mid \bar{Z}_s \right\}
  &~=~ \beta_0^* + \delta_1^*a_s + \delta_2^* \ta_{\tau,s} + \E\{h(\bXcN)\mid \bar{Z}_s \}  \\
  &~=~  \E \left\{ Y_s(\ba) \mid  \bar{Z}_s \right\}  \\
  &~=~ \E \left\{ Y_s(\ba) \mid 
    \bA = \ba, \bar{Z}_s \right\} \\
&~=~ \E \left( Y_s \mid  \bA = \ba, \bar{Z}_s \right) \\
&~=~ \E \left( \beta_0 + \delta_1A_s + \delta_2 \tA_{\tau,s} \mid  
\bA = \ba, \bar{Z}_s \right),
\end{split}\label{fullIdentification}
\end{align}
we must have $\delta_1=\delta_1^*$ and $\delta_2=\delta_2^*$. We use splines to allow for an arbitrary form of dependence between $\bar{Z}$ and $Y$, and include them directly in the regression:
\begin{align}
  \begin{split}
  \E \left( \beta_0 + \delta_1A_s + \delta_2 \tA_{\tau,s} \mid  
  \bA = \ba, \bar{Z}_s \right)
~\approx~ \beta_0 + \delta_1A_s + \delta_2 \tA_{\tau,s} + \spl(\bar{Z}_s) \\
~\approx~ \beta_0 + \delta_1A_s + \delta_2 \tA_{\tau,s} + \spl_1(Z_s^{(1)}) + \cdots + \spl_q(Z_s^{(K)}).
\end{split}\label{splineIdentification}
\end{align}
The second line of (\ref{splineIdentification}) implicitly assumes that the spline components enter additively, an assumption which can be tested. In the presence of non-additivity, a tensor product of the components should be used which allows for general interactions at considerable computational cost \citep{wood2006low}.

\section{Bayesian inference and computational algorithm}\label{s:estimation}

The identification results (\ref{fullIdentification}) and (\ref{splineIdentification}) allow unbiased estimation of $\delta_1$ and $\delta_2$ using a regression of the observed response onto the direct and spill-over treatments as well as the spline estimates of $\bar{Z}_s(\tau)$.  Implementing this involves three steps:  Step 1 parametrizes and estimates the propensities $g_{\tau,s}$ of direct and spill-over treatment. Step 2 estimates a preliminary posterior for the range parameter $\tau$, which must be done in a separate step for reasons discussed below. Step 3 estimates final posterior distributions for all parameters in (\ref{eq:estimated}) via Markov chain Monte Carlo sampling. 

The propensities of direct treatment that are tackled in Step 1 are first estimated by regressing $A_s$ onto $X$. 
This requires parametrizing the form of $f(A)$, and identifying a correctly specified propensity score. The form of this score can vary in complexity. The simplest case is that of a local treatment assignment mechanism, i.e., the distribution of $A_s$ is influenced by $X_s$ only.  This would simply entail a regression on local covariates. A moderately complex case would allow for nearby $X$ to inform the propensity of treatment. A very general case would allow $\bA$ to be spatially-dependent, conditional on $\bX$. That is, $A_s$ would depend directly on nearby $A$. 

Estimating the spill-over propensity component of Step 1 is similar. First, a family of parametric distributions must be identified. One intuitive method of doing this is to select several candidate distributions based on the form of $A_s$, and select among them by simulating values of $\tA_s$. For example, if $A_s$ is binary, then the potential candidates for the distribution of $\tA$ must be nonnegative and allow for point mass at zero. Obvious contenders are zero-inflated lognormal and zero-inflated Gamma distributions. A natural way to select between them is to simulate from the estimated propensities of $A_s$, to get simulated $\tA_s$ values using different reasonable $\tau$. The empirical distributions of these simulated $\tA_s$ will often suggest one family of distributions. With a chosen distribution in hand, the parameters $\bar{Z}$ at each location can be estimated directly from the field $\bX$ and $\tau$. Because these parameters will be conditioned on by entering into a splined regression, it is advantageous that their values have reasonable spread. To this end, one-to-one transformations of the parameters such as log and logit are helpful.

Step 2 involves identifying a plausible set of $\tau$ values to be used in Step 3. Because $\bar{Z}(\tau)$ represents a propensity score, estimating $\tau$ directly in the final model is problematic. It is clear from the definition of a propensity score that the response $Y$ should not provide any information on the propensity of treatment. However, estimating a response model such as (\ref{eq:cutReg}) which includes $\bar{Z}(\tau)$ directly does just that, since $Y$ can influence $\bar{Z}(\tau)$ through $\tau$. This problem is articulated in \cite{mccandless2010cutting, saarela2015bayesian, saarela2016bayesian, zigler2013model}; and  \cite{zigler2016central}. While steps can be taken to mitigate feedback from $Y$ to $\bar{Z}$ issues remain. 

Our solution to this issue takes inspiration from the standard two-step propensity score treatment in which propensity scores are first estimated and treated as fixed, and then conditioned on in an outcome model. Because $\tau$ is unknown, estimating $\bar{Z}(\tau)$ in advance is impossible. However, estimating the  model with feedback in (\ref{eq:cutReg}) does give approximate estimates of $\tau$. From this approximate posterior of $\tau$, a set of reasonable $\tau$ values $(\tau_1, \ldots, \tau_T)$ covering the plausible range of $\tau$ can be identified. Then $\bar{Z}(\tau_1), \ldots, \bar{Z}(\tau_T)$ can be pre-computed and conditioned on simultaneously in the response model in Step 3. Because each of these $\bar{Z}(\tau_t)$ are computed before the response model, the feedback issue is resolved.

Therefore in Step 2 we estimate
\begin{align}    
Y_s &= \beta_0 + \delta_1 A_s + \delta_2 \tA_{\tau,s} + \sum_{k=1}^K \spl_k (Z_s^{(k)}(\tau)) + \epsilon_s.
\label{eq:cutReg}
\end{align}
where $\epsilon_s$ is distributed independent Normal$(0, \sigma^2)$. An attempt to cut the feedback from $Y$ to $\bar{Z}$ is made by estimating $\tau$ in the Metropolis step using only $\tA_\tau$ while holding $\bar{Z}(\tau)$ fixed. A recommended plausible set for $\tau$ might then be $\{\hat{\tau}, \hat{\tau} \pm 2s, \hat{\tau} \pm 4s \}$, where $\hat{\tau}$ and $s$ are the posterior mean and standard deviation of $\tau$ in (\ref{eq:cutReg}). 

Finally in Step 3 each fixed $\bar{Z}(\tau_t)$ enters the final model as
\begin{align}
Y_s &= \beta_0 + \delta_1 A_s + \delta_2 \tA_{\tau,s} + \sum_{t=1}^T \sum_{k=1}^K \spl_{tk} (Z_s^{(k)}(\tau_t)) + \epsilon_s.
\label{eq:estimated}
\end{align}
This model produces accurate posteriors on all variables. Although each each $\tau_t$ is fixed within the $\bar{Z}$ terms, $\tau$ can still vary within $\tA_{\tau,s}$. For the spline terms in (\ref{eq:cutReg})--(\ref{eq:estimated}), we use B-spline expansions taken at fixed intervals over the variables' range of values \citep{eilers1996flexible, ngo2004smoothing}. All regression coefficients are estimated using Gibbs sampling; $\tau$, which now enters only through $\tA_\tau$, uses a Metropolis step.  If Assumptions \ref{unconfoundedness}--\ref{parametricDensity} hold, we recover unbiased estimate of the treatment effects. Comparing the forms of the assumed true model (\ref{eq:responseForm}) and the estimated model (\ref{eq:estimated}) shows that we have essentially replaced the unknown $h(\bXcN)$ with flexible functions of $\bar{Z}$.

\section{Simulation study}\label{s:simstudy}

We examine the performance of this method using simulated data, which take inspiration from the wildfire/air pollution data in Section 7. Since we use a binary treatment in Section \ref{s:fires} to indicate the presence of a fire, we use $A_s \in \{0,1\}$ here. In addition, we assume $A_s$ at different locations is independent conditional on local $X_s$. This precludes the more complex cases of independence conditional on $\bX$ or conditional dependence. Doing this allows for more straightforward modeling of $g_{\tau,s}$, as shown in \ref{ss:estimation}.

We generate the data as follows. The fields $\bX$, $\bA$, and $\bY$ are generated on $n^{1/2} \times n^{1/2}$ grids, with $n=25, 100$ on the unit square $[0,1]\times[0,1]$. We generate $N=100$ independent repeated observations of the fields for each dataset. Thus each complete dataset involves $n\times N$ different data points. The single covariate $X_{s} \in \mathbb{R}^1$ is a mean zero, variance one, Gaussian process and with isotropic exponential covariance and spatial range 0.6. The binary direct treatment $A_{s}$ is determined locally and distributed independently $\text{Bernoulli}\left\{\text{expit}(X_{s} - 3)\right\}$. The continuous spill-over treatment takes the form $\tA_{\tau,s}=\sum_{s'} \omega_\tau \left(\| s-s' \| \right)A_{s'}$, with $\omega_\tau$ a Gaussian kernel as defined in Example \ref{ex:GK} and $\tau=0.3$. Several versions of the confounder $h(\bX)_{s}$ are generated as follows: a weighted average $W_s$ is taken of the $\bX$ values using a Gaussian kernel with $\tau = 0.5$ and weights normalized to sum to 1. Simulations are run with $h(\bX)$ set to $W_s$, $-(W_s)^3$, and $\exp(W_s)$. Lastly, $Y_{s}$ follows the form of (\ref{eq:responseForm}), with $\beta_0=0$, $\delta_1 = \delta_2 = 1$, and $\eps_{s}$ independently distributed standard normal. Each setting is repeated 500 times.

\subsection{Estimation} \label{ss:estimation}

Following the three steps outlined in Section \ref{s:estimation}, we first parametrize and estimate $g_{\tau,s}$. Because $A_s$ is assumed to be conditionally independent given $X_s$, we can estimate $Z$ components for the distributions of $A$ and $\tA$ separately. $A_s$ is binary, so we assume it has a Bernoulli distribution with the correctly specified propensity in which $\logit\{\E (A_s)\}$ is affine in $X_s$. Its distribution is then captured with the standard propensity score $Z_s^{(1)} = \Prob (A_s=1 \mid X_s)$. These values can be estimated with a simple logistic regression from $A_s$ onto $X_s$, with $Z_s^{(1)}$ set to the log of the fitted values. 

We then we choose a parametric form for the distribution of $\tA_{\tau,s}$. From our estimated $Z_s^{(1)}$, we use different plausible $\tau$ values to generate simulated A, which we then use to get an empirical distribution of simulated $\tA$. Examination of these distributions leads us to choose a zero-inflated lognormal distribution for $\tA_s$:
\begin{align*}
  \Prob(\tA = 0 \mid X_s) = p_0, \qquad 
  \Prob(\tA = v \mid  \tA > 0, X_s) = \frac{1}{v \sigma \sqrt{2\pi}} \exp \left\{ -\frac{(\log v - \mu)^2}{2 \sigma^2} \right\}.
\end{align*}
Rather than use the three parameters $p_0$, $\mu$, and $\sigma^2$ for our $Z^{(2)}_s$, $Z^{(3)}_s$, and $Z^{(4)}_s$, we choose three more stable one-to-one transformations: $\logit(p_0)$, $\log \{ \E (\tA) \}$, and $\log\{\Var (\tA)\}$. 

In place of Step 2 the $(\tau_1, \ldots, \tau_T)$ values used are $\{0.25, 0.35, 0.45, 0.55\}$, which surround but do not contain the true $\tau = 0.3$. Rather than re-estimate these values with each simulation repetition, we use this set to ensure comparability across repetitions. Finally, Step 3 uses Markov chain Monte Carlo to estimate all variables in (\ref{eq:estimated}). Further estimation details are provided in Appendix \ref{a:MCMC}.

In addition to the proposed generalized propensity score model, we estimate three comparison models: (i) the oracle model 
[$\E(Y_s) = A_s + \tA_{\tau,s} + h(\bX)_s$] 
is the true model which includes otherwise unknown $h(X_\cD)$ as a covariate, (ii) the local only model 
[$\E(Y_s) = A_s + \tA_{\tau,s} + \sum_j \spl_j (X_s^j) $] 
conditions on local covariates using splines, and (iii) the naive model 
[$\E(Y_s) = A_s + \tA_{\tau,s}$] 
simply regresses the outcome onto the treatments, but does not incorporate any causal conditioning.

\subsection{Simulation results}

Tables \ref{t:nn10:bias}  and \ref{t:nn10:cvg} show the simulation bias and coverage for the $10 \times 10$ grids. The Naive model does very poorly in all scenarios, indicating substantial confounding between $A$ and $Y$. The generalized propensity score model performs substantially better than both the Local only and the Naive models, although, intuitively, the Local only model does show reasonable direct effect estimates. In most cases, the generalized propensity score model performs comparably to the Oracle model. Results for the $5 \times 5$ grids are similar. 

\begin{table}[h!]
\small
\centering
\caption{Simulation bias for $10 \times 10$ grids multiplied by 1,000, with standard errors}
  \label{t:nn10:bias}
\begin{tabular}{clccc}
\\
$h(X_{\mathcal{D}})_s$ & Model &
$\delta_1$ & $\delta_2$ & $\tau$ \\
\\
\multirow{4}{*}{$\displaystyle W_s$ } 
& Oracle 
& \makecell{ 0.2   \scriptsize (1.8)  }
& \makecell{ -0.7   \scriptsize (0.9)  }
& \makecell{ 0.1   \scriptsize (0.2)  } \\
& Generalized propensity score 
& \makecell{ 1.4   \scriptsize (1.9)  }
& \makecell{ 0.5   \scriptsize (1)  }
& \makecell{ 0   \scriptsize (0.2)  } \\
& Local Only 
& \makecell{ 2.6   \scriptsize (2)  }
& \makecell{ -72.1   \scriptsize (1.1)  }
& \makecell{ 69.1   \scriptsize (0.4)  } \\
& Naive 
& \makecell{ 236.7   \scriptsize (1.9)  }
& \makecell{ 52.7   \scriptsize (1.5)  }
& \makecell{ 79.2   \scriptsize (0.5)  } \\
\\
\multirow{4}{*}{ $- \left(  \displaystyle W_s \right)^3$ } 
& Oracle 
& \makecell{ 0.1   \scriptsize (1.8)  }
& \makecell{ -0.7   \scriptsize (0.9)  }
& \makecell{ 0.1   \scriptsize (0.2)  } \\
& Generalized propensity score 
& \makecell{ 1.1   \scriptsize (1.9)  }
& \makecell{ -0.1   \scriptsize (1)  }
& \makecell{ 0.1   \scriptsize (0.2)  } \\
& Local Only 
& \makecell{ 1   \scriptsize (2.1)  }
& \makecell{ 28.5   \scriptsize (1.3)  }
& \makecell{ -39.7   \scriptsize (0.4)  } \\
& Naive 
& \makecell{ -205.9   \scriptsize (2.8)  }
& \makecell{ -104.1   \scriptsize (1.6)  }
& \makecell{ -53.1   \scriptsize (0.6)  } \\
\\
\multirow{4}{*}{ $\exp \left( \displaystyle W_s \right)$ } 
& Oracle 
& \makecell{ 0.3   \scriptsize (1.8)  }
& \makecell{ -0.6   \scriptsize (0.9)  }
& \makecell{ 0.1   \scriptsize (0.2)  } \\
& Generalized propensity score 
& \makecell{ 1.5   \scriptsize (1.9)  }
& \makecell{ 0.9   \scriptsize (1.1)  }
& \makecell{ 0.1   \scriptsize (0.3)  } \\
& Local Only 
& \makecell{ 2.8   \scriptsize (2.3)  }
& \makecell{ -101.4   \scriptsize (2.1)  }
& \makecell{ 120.7   \scriptsize (1.4)  } \\
& Naive 
& \makecell{ 381.3   \scriptsize (2.8)  }
& \makecell{ 105.3   \scriptsize (2.3)  }
& \makecell{ 114.2   \scriptsize (1.3)  } 
\end{tabular}
\end{table}

\begin{table}[h!]
\small
\centering
\caption{Simulation coverage for $10 \times 10$ grids, with standard errors}
\label{t:nn10:cvg}
\begin{tabular}{clccc}
\\
$h(X_{\mathcal{D}})_s$ & Model &
$\delta_1$ &
$\delta_2$ &
$\tau$  \\
\\
\multirow{4}{*}{$\displaystyle W_s$ } 
& Oracle 
& \makecell{ 95 \scriptsize (1)  }
& \makecell{ 94.6 \scriptsize (1)  }
& \makecell{ 93.6 \scriptsize (1.1)  } \\
& Generalized propensity score 
& \makecell{ 93.8 \scriptsize (1.1)  }
& \makecell{ 94.4 \scriptsize (1)  }
& \makecell{ 93.6 \scriptsize (1.1)  } \\
& Local Only 
& \makecell{ 93.2 \scriptsize (1.1)  }
& \makecell{ 8.8 \scriptsize (1.3)  }
& \makecell{ 0 \scriptsize (0)  } \\
& Naive 
& \makecell{ 0 \scriptsize (0)  }
& \makecell{ 27.6 \scriptsize (2)  }
& \makecell{ 0 \scriptsize (0)  }  \\
\\
\multirow{4}{*}{ $- \left(  \displaystyle W_s \right)^3$ } 
& Oracle 
& \makecell{ 95.2 \scriptsize (1)  }
& \makecell{ 94.6 \scriptsize (1)  }
& \makecell{ 93.8 \scriptsize (1.1)  } \\
& Generalized propensity score 
& \makecell{ 95.2 \scriptsize (1)  }
& \makecell{ 93.8 \scriptsize (1.1)  }
& \makecell{ 92.8 \scriptsize (1.2)  } \\
& Local Only 
& \makecell{ 93.8 \scriptsize (1.1)  }
& \makecell{ 77.2 \scriptsize (1.9)  }
& \makecell{ 0 \scriptsize (0)  } \\
& Naive 
& \makecell{ 2 \scriptsize (0.6)  }
& \makecell{ 6 \scriptsize (1.1)  }
& \makecell{ 0 \scriptsize (0)  } \\
\\
\multirow{4}{*}{ $\exp \left(  \displaystyle W_s \right)$ } 
& Oracle 
& \makecell{ 95.4 \scriptsize (0.9)  }
& \makecell{ 94.2 \scriptsize (1)  }
& \makecell{ 93.8 \scriptsize (1.1)  } \\
& Generalized propensity score 
& \makecell{ 93 \scriptsize (1.1)  }
& \makecell{ 90.2 \scriptsize (1.3)  }
& \makecell{ 90.2 \scriptsize (1.3)  } \\
& Local Only 
& \makecell{ 90.8 \scriptsize (1.3)  }
& \makecell{ 4.6 \scriptsize (0.9)  }
& \makecell{ 0 \scriptsize (0)  } \\
& Naive 
& \makecell{ 0 \scriptsize (0)  }
& \makecell{ 8.2 \scriptsize (1.2)  }
& \makecell{ 0 \scriptsize (0)  }
\end{tabular}
\end{table}

\section{Estimating the causal effect of wildland fires on air pollution}\label{s:fires}

Wildland fires release harmful particles and gasses impacting air quality near the fire and downwind \citep{larsen2018impacts}. Fine particulate matter smaller than 2.5 $\mu$m (PM$_{2.5}$) have been linked to adverse cardiorespiratory health outcomes \citep{brook2007air,dominici2006fine,corrigan2018fine,rappold2012cardio,weber2016assessing}. For these reasons, understanding the causal effect of wildland fires on air pollution across space is of significant interest. Specifically, we are interested in the time-averaged causal effect of wildfires on ambient PM$_{2.5}$ concentrations across Western United States from 2005 to 2018. 

\subsection{Data}
\begin{figure} 
\begin{subfigure}{.5\textwidth} 
  \centering 
  \includegraphics[width=1.05\linewidth]{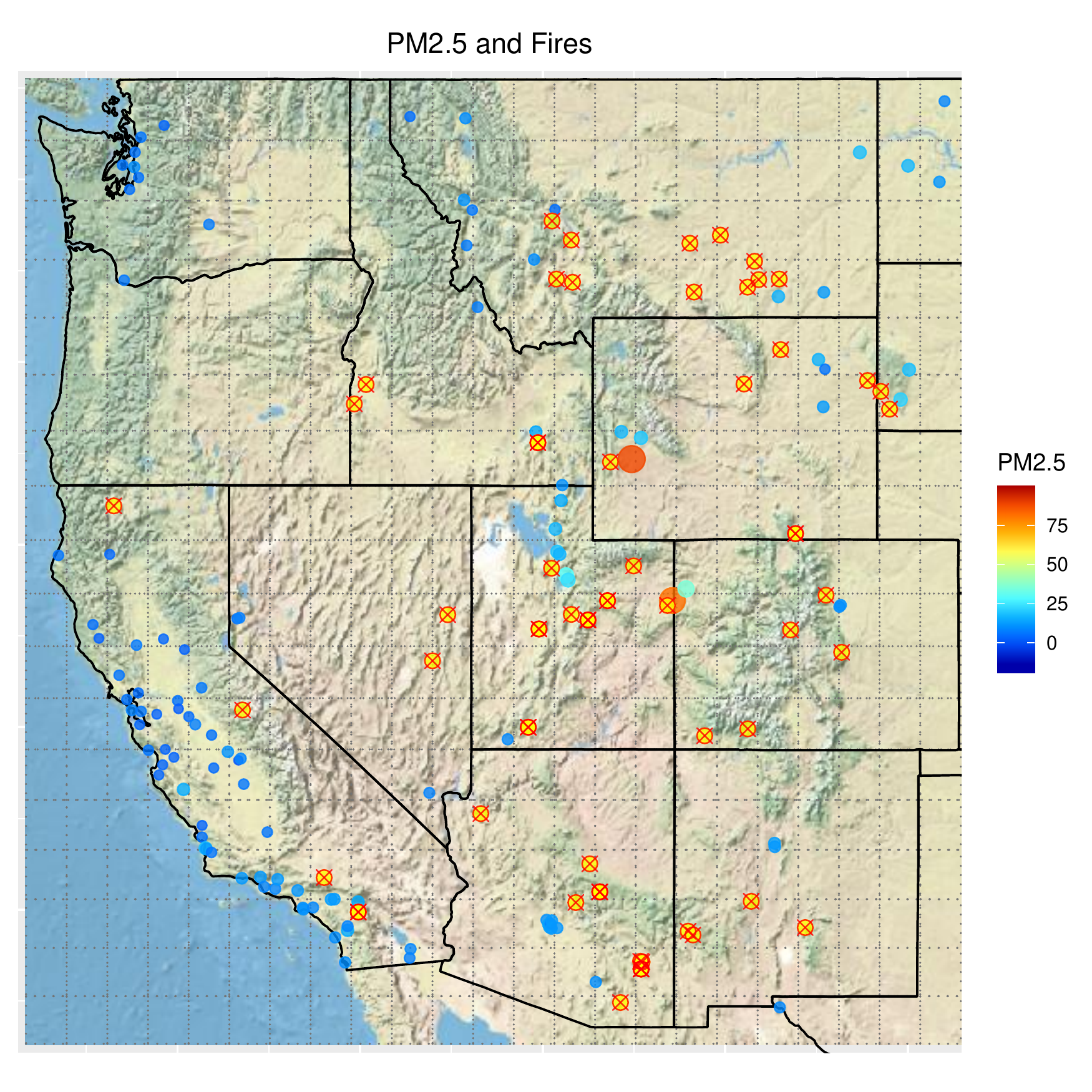}
  \caption{Fires ($A$) and PM$_{2.5}$ ($Y$)}
\end{subfigure}
\begin{subfigure}{.5\textwidth}
  \centering
  \includegraphics[width=.9\linewidth]{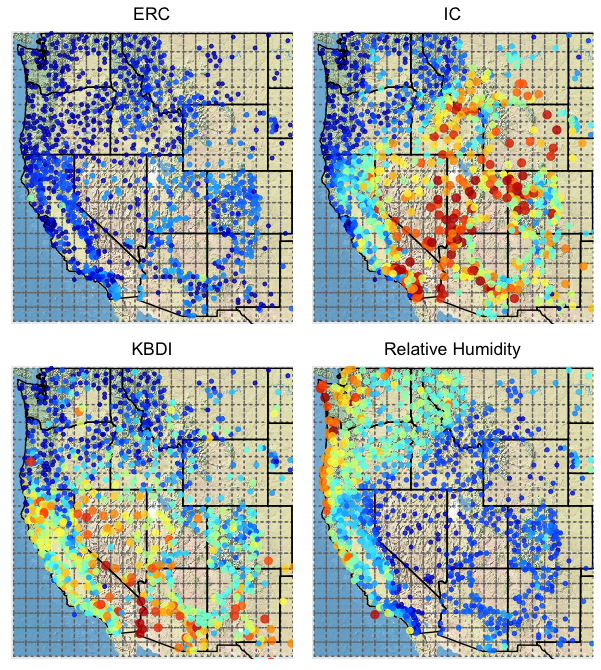} 
  \caption{Four of the 10 covariates $X^j$}
\end{subfigure}
\caption{Data snapshot on July 1, 2012. Energy Release Component (ERC) and Ignition Component (IC) are two of National Fire Danger Rating System Components;  KBDI refers to the Keetch-Byram drought index. In (a) fires are shown as cross-hatched circles and PM$_{2.5}$ locations are shown as solid circles} 
\label{fig:data}
\end{figure}

The response $Y$ consists of 24-hour average PM$_{2.5}$ concentrations  measured in $\mu$g/m$^3$ at 416 
measurement sites, some of which are plotted in Fig.~\ref{fig:data}. Observations are collected every one, three or six days depending on the station. The data are publicly available and provided by the Environmental Protection Agency. For each location, the long-term mean is subtracted.

The dates and locations of fires are compiled from a mix of satellite data and incident reports reported to the Geospatial Multi-Agency Coordination Wildland Fire Support program. Because the focus of our analysis is on PM$_{2.5}$ only fires larger than 1,000 acres are included in the analysis. Among the 3,930 fires, 34.8\% of fires are missing either a start or end date. For these fires we impute missing values by modeling fire duration as a linear function of log(area burned). 

Lastly, 11 confounders $X^1, \ldots, X^{11}$ are included in the treatment balancing score. These include the four components of the National Fire Danger Rating System (energy release component, burning index, ignition component, and spread index) which are used to monitor daily risk of fire in the United States. The other variables used in the balancing score include elevation, daily temperature, relative humidity, wind speed, precipitation level, the Keetch-Byram drought index, and the numeric day of the year.  A snapshot of the treatment, response, as well as the energy release component, ignition component, Keetch-Byram drought index, and relative humidity for one day are shown in Figure \ref{fig:data}. 

Our analysis treats each daily air observation as the center of a $9 \times 9$ grid, with a height and width of 9 degrees latitude/longitude. For each such grid, only the center grid cell has a response $Y_s$ value. However, all 81 grid cells have covariates $X_s^{j}$ and direct treatment $A_s$ values. Each grid cell receives direct treatment $A_s=1$ if there was at least one fire in the cell on that particular day; 0 otherwise. Each $X_s^{j}$ value is taken to be the mean of the observed covariates in each cell/day combination. For cell/days with no observed values, a value is imputed from nearby cells using a kernel smoother as implemented in the ``fields'' R package \citep{nychka2014fields}. The end result is 605,414 observed grids, each of which contain $9 \times 9$ grids for $A_\cD$ and $X^j_\cD, j=1,\ldots, 11$, as well as a centered $Y_s$ value. Finally, any grid cells whose centers extend outside of the Western United States are disregarded and excluded from analysis. In this context, the direct effect of treatment consists of the causal effect on $Y_s$ from a fire in the same grid cell ($A_s = 1$), whereas the indirect effect consists of the causal effect on $Y_s$ from $A_{s'}$ in other cells ($s \ne s'$). As in the simulation study, each of these grids are treated as independent. In addition to the generalized propensity score model, we estimate a model that conditions on the local covariates only, using splines. 

We use the same form of $g_{\tau,s}$ as given in Section \ref{s:simstudy}. $A_s$ at different locations are assumed to be conditionally independent given $X_s$, which allows us to estimate separate components for $A_s$ and $\tA_{\tau, s}$. The propensity component $\logit\{ \E (A_s)\}$ is estimated as a linear model of 5-element B-splines of $X_s^1, \ldots, X_s^{11}$, and the propensity of $\tA$ is assumed to be zero-inflated lognormal. Conditioning on local $X_s$ only is justified because we posit that it is the local $X_s$ that contains the vast majority of information about the propensity of fire, with locations further away giving far less information. 

\subsection{Results}

Table \ref{t:results} shows the results. The causal direct effect estimate given by the generalized propensity score model is  1.03 $\mu$g/m$^3$ of PM$_{2.5}$, or 11.9\% of the annual mean PM$_{2.5}$ observed throughout. The range parameter $\tau$ is estimated to be 1.53 degrees of latitude/longitude, suggesting that fires impact up to roughly 3 degrees away. The estimate of 0.13 for $\delta_2$ represents the height of spill-over kernel at its peak. All of $\delta_1$, $\delta_2$, and $\tau$ are highly significant. The estimated direct effect from the local-only model is 12\% larger than the estimate from the generalized propensity score model, and the local-only model has an implausibly large and imprecise estimate of $\tau$. 

\begin{figure} 
\centering 
  \includegraphics[width=.8\linewidth]{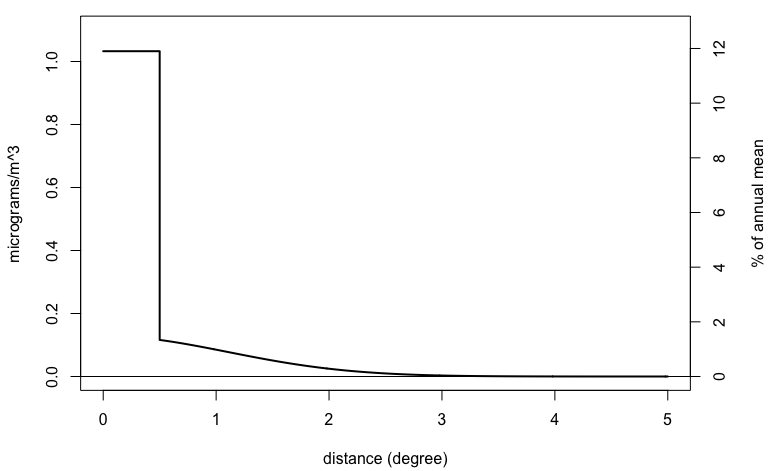}
  \caption{Causal effect of a fire on PM$_{2.5}$ by distance, as measured in degrees of latitude/longitude.  The left axis shows the raw causal increase in PM$_{2.5}$; the right axis shows this as a percentage of annual mean PM$_{2.5}$ levels.}
\label{fig:results}
\end{figure}

Figure \ref{fig:results} illustrates the implied causal effect of fire at different distances from the generalized propensity score model. Taking the center of a grid cell as our vantage point, the direct effect of one or more fires in the same grid cell has a time-averaged causal increase of 1.03 $\mu$g/m$^3$ of PM$_{2.5}$, which corresponds to the step from 0 to 0.5 in the east/west or north/south direction; slightly more than 0.5 when at an angle. As the fire gets progressively further away, the causal effect decays smoothly until it approaches 0 roughly 3 grid cells away. Intuitively this kernel extending out from 0 is completely determined by $\tau$ and $\delta_2$: $\tau$ corresponds to the width of the kernel; $\delta_2$ is the height of the kernel at its peak. 

The wildfire analysis makes several simplifications that are important to consider. First, treating $A_s$ as binary sacrifices information on the number and size of fires in a given grid cell. Extending this method to incorporate information on the size of the fire would preserve information. Moreover, we assume $\tau$, $\delta_1$, and $\delta_2$ are fixed, although it is possible that they naturally vary across different fires and locations. However, there is not enough information in the data to identify these differences. Additionally, we do not consider time-varying effects, as we focused on the contribution to time-averaged PM$_{2.5}$ levels. Another important simplification is the treatment of separate days as independent. There are temporal trends in the treatment, response, and covariates, and our assumption of independence may inflate the amount of information that our data appear to have. 

\begin{center}
\begin{table}
\caption{Posterior mean (95\% Credible Interval)}~~
\label{t:results}
\small
\centering
\begin{tabular}{rccc}
& Direct Effect ($\delta_1$) & Spillover Effect ($\delta_2$) & Bandwidth ($\tau$)  \\ \\
Local Only & 
1.15 (1.05,1.25) & 0.11 (0.09, 0.12) & 17.37 (8.28, 42.12) \\ \\
Generalized propensity score & 
1.03 (0.93,1.14) & 0.13 (0.03, 0.25) & 1.53 (1.17, 2.88)
\end{tabular}
\end{table}
\end{center}

\section{Discussion}\label{s:discussion}

The generalized propensity score method presented here establishes a new framework to recover causal direct and spill-over effects in the presence of spatial interference. The inherent dimensionality issues of the problem are dealt with via a novel propensity score approach, which uses a Bayesian spline-based regression model and a dimension reduction approximation to make the problem tractable.
However, there are several critical yet strong assumptions that must hold for our method to perform well. The method hinges on a correctly specified propensity score $g_{\tau,s}(\bX)$ as well as a correctly specified potential outcomes model in (\ref{eq:responseForm}). This includes accommodating conditionally dependent $\bA$, and correctly characterizing the spatial dependence on $A_s$ from nearby $X$.  Moreover, the no unmeasured confounders assumption is always a strong, but necessary, assumption for causal inference on observational data. In practice considerable effort should be made to include any potential confounders for this reason. Lastly, we rely crucially on the assumption that the distribution of treatments $(A_s, \tA_s)$ can be encapsulated with the parameters $\bar{Z}_s$ of the propensity score $g_{\tau,s}$. This will rarely be completely accurate in practice, so effort should be made to select an appropriate parametric form for $g_{\tau,s}$.

\appendix

\section{Proof of Theorem 1} \label{a:proof}

Claim 1: $g_\tau$ is a balancing score.
\begin{proof}
By the definition of a propensity score, $g_{\tau,s}(X_\cD)$ has the property that $\Prob \{(A_s=\eta,\tA_s=\nu) \mid  X_\cD, g_{\tau,s}\} = \Prob \{(A_s=\eta,\tA_s=\nu) \mid  X_\cD \}$ which implies ~$X_\cD \ind (A_s,\tA_s) \mid  g_{\tau,s}$.   
And thus $g_{\tau,s}$ is a balancing score for our covariates $\bXcN$.  
As noted by \cite{hirano2004propensity} this balancing is a characteristic of $g_{\tau,s}$, and does not rely on any unconfoundedness in the response yet.
\end{proof}

Claim 2: for all levels $\nu$, 
\begin{align*}
\Prob \left[ A_s=\eta, \tA_{\tau,s}=\nu \mid  Y_s \{ a_s = \eta, \ta_s = \nu \}, ~g_{\tau,s}(\eta,\nu,\bX)\right] = \Prob \{ A_s=\eta, \tA_s=\nu \mid g_{\tau,s}(\eta,\nu,\bXcN) \}.	
\end{align*}
(To ease notation, now let $\bar{A}_s = (A_s, \tA_{\tau,s})$ and $\bar{a}_s = (a_s, \ta_{\tau,s})$.)

\begin{proof}
We can then write
\begin{align*}
\Prob \left\{ \bar{A}_s=(\eta,\nu) \mid  g_{\tau,s}(\eta,\nu,\bXcN)\right\} &= f_{\bar{A}_s}\left\{ \eta, \nu \mid g_{\tau,s}(\eta,\nu,\bXcN)\right\}\\
  &= \int f_{\bar{A}_s} \left\{ \eta, \nu \mid \bX,g_{\tau,s}(\eta,\nu,\bXcN)\right\}
    \dd F_{\bXcN} \left\{ \bX\mid g_{\tau,s}(\eta,\nu,\bXcN)\right\} \\  
  &= \int f_{\bar{A}_s} (\eta, \nu  \mid  \bX)
  \dd F_{\bXcN} \left\{ \bX \mid  g_{\tau,s}(\eta,\nu,\bXcN)\right\}\\
  &= \int g_{\tau,s}(\eta,\nu, \bX) \dd F_{\bXcN}\left\{ \bX \mid  g_{\tau,s}(\eta,\nu,\bXcN)\right\} \\
  &= g_{\tau,s}(\eta,\nu,\bXcN),
\end{align*}
\begin{align*}
  \Prob \big[ \bar{A}_s = (\eta, \nu)~&\mid ~
        Y_s\big\{ \bar{a}_s = (\eta, \nu)\big\}, ~g_{\tau,s}(\eta,\nu,\bXcN) \big] \\ 
  &= f_{\bar{A}_s}\left[ \nu\mid g_{\tau,s}(\eta,\nu,\bXcN), Y_s\left\{ \bar{a}_s =(\eta, \nu) \right\} \right]\\
  &= \int f_{\bar{A}_s} \left[ \eta, \nu  \mid  \bX, g_{\tau,s}(\eta,\nu,\bXcN),  Y_s \left\{ \bar{a}_s =(\eta,\nu) \right\}\right]
    \dd F_{\bXcN} \left[ \bX \mid   Y_s\left\{ \bar{a}_s =(\eta, \nu) \right\}, g_{\tau,s}(\eta,\nu,\bXcN)\right] \\
  &= \int f_{\bar{A}_s} (\eta, \nu  \mid  \bX)
    \dd F_{\bXcN} \left[ \bX \mid   Y_s\left\{ \bar{a}_s =(\eta, \nu) \right\}, g_{\tau,s}(\eta,\nu,\bXcN)\right] \\
  &= \int g_{\tau,s}(\eta,\nu, \bX)
    \dd F_{\bXcN}\left[\bX \mid   Y_s\left\{ \bar{a}_s =(\eta, \nu) \right\}, g_{\tau,s}(\eta,\nu,\bXcN)\right] \\
  &= g_{\tau,s}(\eta,\nu,\bXcN).
\end{align*}
Combining these gives Claim 2, which then implies our result.
\end{proof}

\section{Bayesian estimation details for simulation}\label{a:MCMC}

Uninformative priors are used for all parameters except $\tau$ which receives a mildly informative prior. Markov chain Monte Carlo iterations begin at maximum likelihood values for all parameters except $\tau$, which requires an initial estimate. A burn-in length of 7,500 iterations is used, after which we sample 22,500 iterations. Gibbs sampling is used for all parameters except $\tau$, which we transform and sample using Metropolis sampling, with an adaptive tuning scheme during the burn-in. Specifically, we use a normal proposal distribution for $\log(\tau - \frac{1}{d})$, where $d$ is the number of grid cells along each axis. This prevents the $\tau$ samples from becoming pathologically small, in which case the kernel cannot reach the neighboring cells and $\delta_2$ becomes arbitrary large. The comparison models are estimated with similar parameter settings. 

For convenience, define $\beta$ as the vector of $\beta_0$, $\delta_1$, $\delta_2$, and the spline coefficients; let $\mu_{s} = \beta_0 + \delta_1 A_s + \delta_2 \tA_s + \sum_{j=1}^J b_j^{(0)} B_j^{(0)}(e_s) + \sum_{t=1}^T \sum_{k=1}^q  \sum_{j=1}^J b_{j,t}^{(k)} B_{j,t}^{(k)} \{ Z_s^{(k)}(\tau_t) \}$; let $M$ be the matrix with columns $\1_{(nN)}$, $\bA^{vec}$, $\tA(\tau)^{vec}$, and the B-splines bases; and let $\Sigma_0 = diag(1000, \ldots ,1000)$. We then specify
\begin{align*} 
Y_s \mid  \tau, \bbeta, \sigma^2_\eps &\sim \text{Normal} 
\{ \mu_{s}(\tau,\bbeta),~\sigma^2_\eps I \}   \\ 
\tl &\sim \text{Normal}(-1,~1) \\
\sigma^2_\eps &\sim \text{InverseGamma}(0.001,~ 0.001) \\
\bbeta  &\sim \text{Normal} (0, ~\Sigma_0 ) \\ 
\bbeta  \mid  \tau, \sigma^2_\eps &\sim \text{Normal} \{ (\Sigma_0^{-1} + M^{\Top} M / \sigma^2_\eps)^{-1} M^{\Top} \bY^{vec}/\sigma^2_\eps,~ (\Sigma_0^{-1} + M^{\Top} M / \sigma^2_\eps)^{-1} \} \\
\sigma^2_\eps \mid  \bbeta, \tau &\sim \text{InverseGamma} \{ 0.001 + (nN)/2 ,~ 0.001 + (\bY^{vec}- M\bbeta)^\top (\bY^{vec} - M\bbeta)/2 \} \\
\tl \mid  \bbeta, \sigma^2_\eps &\propto \text{Normal}_{Y}(\mu^{vec},~\sigma^2_\eps I)\times \text{Normal}_{\tl}(0,100).
\end{align*}

\bibliographystyle{plainnat}     
\bibliography{Bibliography}  

\begin{thebibliography}{33}
\providecommand{\natexlab}[1]{#1}
\providecommand{\url}[1]{\texttt{#1}}
\expandafter\ifx\csname urlstyle\endcsname\relax
  \providecommand{\doi}[1]{doi: #1}\else
  \providecommand{\doi}{doi: \begingroup \urlstyle{rm}\Url}\fi

\bibitem[Athey et~al.(2018)Athey, Eckles, and Imbens]{athey2018exact}
Susan Athey, Dean Eckles, and Guido~W Imbens.
\newblock Exact p-values for network interference.
\newblock \emph{Journal of the American Statistical Association}, 113\penalty0
  (521):\penalty0 230--240, 2018.

\bibitem[Bind(2019)]{bind2019causal}
Marie-Ab{\`e}le Bind.
\newblock Causal modeling in environmental health.
\newblock \emph{Annual review of public health}, 40:\penalty0 23--43, 2019.

\bibitem[Brook(2007)]{brook2007air}
Robert~D Brook.
\newblock Is air pollution a cause of cardiovascular disease? updated review
  and controversies.
\newblock \emph{Reviews on environmental health}, 22\penalty0 (2):\penalty0
  115--138, 2007.

\bibitem[Corrigan et~al.(2018)Corrigan, Becker, Neas, Cascio, and
  Rappold]{corrigan2018fine}
Anne~E Corrigan, Michelle~M Becker, Lucas~M Neas, Wayne~E Cascio, and Ana~G
  Rappold.
\newblock Fine particulate matters: the impact of air quality standards on
  cardiovascular mortality.
\newblock \emph{Environmental research}, 161:\penalty0 364--369, 2018.

\bibitem[Cox(1958)]{cox1958planning}
David~Roxbee Cox.
\newblock \emph{Planning of Experiments.}
\newblock Wiley, 1958.

\bibitem[Dominici et~al.(2006)Dominici, Peng, Bell, Pham, McDermott, Zeger, and
  Samet]{dominici2006fine}
Francesca Dominici, Roger~D Peng, Michelle~L Bell, Luu Pham, Aidan McDermott,
  Scott~L Zeger, and Jonathan~M Samet.
\newblock Fine particulate air pollution and hospital admission for
  cardiovascular and respiratory diseases.
\newblock \emph{Jama}, 295\penalty0 (10):\penalty0 1127--1134, 2006.

\bibitem[Eilers and Marx(1996)]{eilers1996flexible}
Paul~HC Eilers and Brian~D Marx.
\newblock Flexible smoothing with b-splines and penalties.
\newblock \emph{Statistical science}, pages 89--102, 1996.

\bibitem[Halloran and Struchiner(1991)]{halloran1991study}
M~Elizabeth Halloran and Claudio~J Struchiner.
\newblock Study designs for dependent happenings.
\newblock \emph{Epidemiology}, pages 331--338, 1991.

\bibitem[Halloran and Struchiner(1995)]{halloran1995causal}
M~Elizabeth Halloran and Claudio~J Struchiner.
\newblock Causal inference in infectious diseases.
\newblock \emph{Epidemiology (Cambridge, Mass.)}, 6\penalty0 (2):\penalty0
  142--151, 1995.

\bibitem[Hirano and Imbens(2004)]{hirano2004propensity}
Keisuke Hirano and Guido~W Imbens.
\newblock The propensity score with continuous treatments.
\newblock \emph{Applied Bayesian modeling and causal inference from
  incomplete-data perspectives}, 226164:\penalty0 73--84, 2004.

\bibitem[Hudgens and Halloran(2008)]{hudgens2008toward}
Michael~G Hudgens and M~Elizabeth Halloran.
\newblock Toward causal inference with interference.
\newblock \emph{Journal of the American Statistical Association}, 103\penalty0
  (482):\penalty0 832--842, 2008.

\bibitem[Larsen et~al.(2018)Larsen, Reich, Ruminski, and
  Rappold]{larsen2018impacts}
Alexandra~E Larsen, Brian~J Reich, Mark Ruminski, and Ana~G Rappold.
\newblock Impacts of fire smoke plumes on regional air quality, 2006--2013.
\newblock \emph{Journal of exposure science \& environmental epidemiology},
  28\penalty0 (4):\penalty0 319, 2018.

\bibitem[Liu and Hudgens(2014)]{liu2014large}
Lan Liu and Michael~G Hudgens.
\newblock Large sample randomization inference of causal effects in the
  presence of interference.
\newblock \emph{Journal of the American Statistical Association}, 109\penalty0
  (505):\penalty0 288--301, 2014.

\bibitem[Manski(2013)]{manski2013identification}
Charles~F Manski.
\newblock Identification of treatment response with social interactions.
\newblock \emph{The Econometrics Journal}, 16\penalty0 (1):\penalty0 S1--S23,
  2013.

\bibitem[McCandless et~al.(2010)McCandless, Douglas, Evans, and
  Smeeth]{mccandless2010cutting}
Lawrence~C McCandless, Ian~J Douglas, Stephen~J Evans, and Liam Smeeth.
\newblock Cutting feedback in bayesian regression adjustment for the propensity
  score.
\newblock \emph{The international journal of biostatistics}, 6\penalty0 (2),
  2010.

\bibitem[Ngo and Wand(2004)]{ngo2004smoothing}
Long Ngo and Matthew~P Wand.
\newblock Smoothing with mixed model software.
\newblock 2004.

\bibitem[Nychka et~al.(2014)Nychka, Furrer, and Sain]{nychka2014fields}
Douglas Nychka, Reinhard Furrer, and S~Sain.
\newblock fields: Tools for spatial data. r package version 7.1.
\newblock \emph{Accessed online}, 10, 2014.

\bibitem[Papadogeorgou et~al.(2019)Papadogeorgou, Mealli, and
  Zigler]{papadogeorgou2019causal}
Georgia Papadogeorgou, Fabrizia Mealli, and Corwin~M Zigler.
\newblock Causal inference with interfering units for cluster and population
  level treatment allocation programs.
\newblock \emph{Biometrics}, 2019.

\bibitem[Perez-Heydrich et~al.(2014)Perez-Heydrich, Hudgens, Halloran, Clemens,
  Ali, and Emch]{perez2014assessing}
Carolina Perez-Heydrich, Michael~G Hudgens, M~Elizabeth Halloran, John~D
  Clemens, Mohammad Ali, and Michael~E Emch.
\newblock Assessing effects of cholera vaccination in the presence of
  interference.
\newblock \emph{Biometrics}, 70\penalty0 (3):\penalty0 731--741, 2014.

\bibitem[Rappold et~al.(2012)Rappold, Cascio, Kilaru, Stone, Neas, Devlin, and
  Diaz-Sanchez]{rappold2012cardio}
Ana~G Rappold, Wayne~E Cascio, Vasu~J Kilaru, Susan~L Stone, Lucas~M Neas,
  Robert~B Devlin, and David Diaz-Sanchez.
\newblock Cardio-respiratory outcomes associated with exposure to wildfire
  smoke are modified by measures of community health.
\newblock \emph{Environmental Health}, 11\penalty0 (1):\penalty0 71, 2012.

\bibitem[Rosenbaum and Rubin(1983)]{rosenbaum1983central}
Paul~R Rosenbaum and Donald~B Rubin.
\newblock The central role of the propensity score in observational studies for
  causal effects.
\newblock \emph{Biometrika}, 70\penalty0 (1):\penalty0 41--55, 1983.

\bibitem[Rubin(1974)]{rubin1974estimating}
Donald~B Rubin.
\newblock Estimating causal effects of treatments in randomized and
  nonrandomized studies.
\newblock \emph{Journal of Educational Psychology}, 66\penalty0 (5):\penalty0
  688, 1974.

\bibitem[Rubin(1980)]{rubin1980randomization}
Donald~B Rubin.
\newblock Randomization analysis of experimental data: The fisher randomization
  test comment.
\newblock \emph{Journal of the American Statistical Association}, 75\penalty0
  (371):\penalty0 591--593, 1980.

\bibitem[Saarela et~al.(2015)Saarela, Stephens, Moodie, and
  Klein]{saarela2015bayesian}
Olli Saarela, David~A Stephens, Erica~EM Moodie, and Marina~B Klein.
\newblock On bayesian estimation of marginal structural models.
\newblock \emph{Biometrics}, 71\penalty0 (2):\penalty0 279--288, 2015.

\bibitem[Saarela et~al.(2016)Saarela, Belzile, and
  Stephens]{saarela2016bayesian}
Olli Saarela, L{\'e}o~R Belzile, and David~A Stephens.
\newblock A bayesian view of doubly robust causal inference.
\newblock \emph{Biometrika}, 103\penalty0 (3):\penalty0 667--681, 2016.

\bibitem[Sobel(2006)]{sobel2006randomized}
Michael~E Sobel.
\newblock What do randomized studies of housing mobility demonstrate? causal
  inference in the face of interference.
\newblock \emph{Journal of the American Statistical Association}, 101\penalty0
  (476):\penalty0 1398--1407, 2006.

\bibitem[Tchetgen and VanderWeele(2012)]{tchetgen2012causal}
Eric J~Tchetgen Tchetgen and Tyler~J VanderWeele.
\newblock On causal inference in the presence of interference.
\newblock \emph{Statistical Methods in Medical Research}, 21\penalty0
  (1):\penalty0 55--75, 2012.

\bibitem[Verbitsky-Savitz and Raudenbush(2012)]{verbitsky2012causal}
Natalya Verbitsky-Savitz and Stephen~W Raudenbush.
\newblock Causal inference under interference in spatial settings: A case study
  evaluating community policing program in chicago.
\newblock \emph{Epidemiologic Methods}, 1\penalty0 (1):\penalty0 107--130,
  2012.

\bibitem[Weber et~al.(2016)Weber, Insaf, Hall, Talbot, and
  Huff]{weber2016assessing}
Stephanie~A Weber, Tabassum~Z Insaf, Eric~S Hall, Thomas~O Talbot, and Amy~K
  Huff.
\newblock Assessing the impact of fine particulate matter (pm2. 5) on
  respiratory-cardiovascular chronic diseases in the new york city metropolitan
  area using hierarchical bayesian model estimates.
\newblock \emph{Environmental research}, 151:\penalty0 399--409, 2016.

\bibitem[Wood(2006)]{wood2006low}
Simon~N Wood.
\newblock Low-rank scale-invariant tensor product smooths for generalized
  additive mixed models.
\newblock \emph{Biometrics}, 62\penalty0 (4):\penalty0 1025--1036, 2006.

\bibitem[Zigler et~al.(2012)Zigler, Dominici, and Wang]{zigler2012estimating}
Corwin~M Zigler, Francesca Dominici, and Yun Wang.
\newblock Estimating causal effects of air quality regulations using principal
  stratification for spatially correlated multivariate intermediate outcomes.
\newblock \emph{Biostatistics}, 13\penalty0 (2):\penalty0 289--302, 2012.

\bibitem[Zigler et~al.(2013)Zigler, Watts, Yeh, Wang, Coull, and
  Dominici]{zigler2013model}
Corwin~M Zigler, Krista Watts, Robert~W Yeh, Yun Wang, Brent~A Coull, and
  Francesca Dominici.
\newblock Model feedback in bayesian propensity score estimation.
\newblock \emph{Biometrics}, 69\penalty0 (1):\penalty0 263--273, 2013.

\bibitem[Zigler(2016)]{zigler2016central}
Corwin~Matthew Zigler.
\newblock The central role of bayes’ theorem for joint estimation of causal
  effects and propensity scores.
\newblock \emph{The American Statistician}, 70\penalty0 (1):\penalty0 47--54,
  2016.

\end{thebibliography}
\end{document}